\documentclass[12pt]{iopart}

\usepackage{graphicx}

\begin{document}
\title{Electromagnetic Field Induced Modification of Branching Ratios for
 Emission in Structured Vacuum}
\author{G. S. Agarwal and Sumanta Das}
\address{Department of Physics, Oklahoma State University,
Stillwater, OK - 74078, USA} \eads{\mailto{agirish@okstate.edu},
\mailto{sumanta.das@okstate.edu}}
\date{\today}

\begin{abstract}
We report a fundamental effect of the electromagnetic field
induced modification of the branching ratios for emission into
several final states. The modifications are especially significant
if the vacuum into which the atom is radiating has a finite
spectral width comparable with the separation of the final states.
This is easily realizable in cavity QED. Further our results are
quite generic and are applicable to any system interacting with a
structured reservoir.
\end{abstract}
\maketitle

Mollow discovered in 1969 how the spectral characteristics of the
radiation emitted by a system are modified quite significantly due
to a coherent field driving the system \cite{mol}. Such spectral
modifications have been the subject of extensive experimental and
theoretical studies \cite{eze} and were explained neatly in terms
of the dressed state
 picture \cite{coh}. The work of Mollow was extended to the case of emission in a
 cavity \cite{wal,mos}. It was further found specially in the context of multilevel systems that
 the driving fields can produce well defined interference minimum in the spectrum
 \cite{zu,gsa,kei,pat}. Such minimum is usually interpreted in terms of the interferences produced
 by different dressed state emissions. Very often this interferences is also referred to as
 the quenching of spontaneous emission \cite{scu}. A related question is-- what is
 the effect of driving fields on branching ratios in emission to multiple states. In a different
 class of experiments Suckewer and co-workers \cite{suk} found definite evidence of external
 field induced changes in branching ratios. Their experimental finding has been rather
 difficult to explain due to complicated nature of the laser plasma used in the experiment.
  It is therefore desirable to look for simpler systems where one can analyze how
  external fields could affect branching ratios.\\

  In this paper we analyze a cavity QED system to highlight the field induced modification
  of the branching ratios. In view of the enormous progress made in the context of cavity
  QED \cite{har,walt,kim} such findings are within the reach of present experiments. Our
  analysis also suggests that change in the branching ratios are notable if one works
  in a regime where separation between the two final states is more than the width
  of the vacuum into which the system is radiating. Thus some of the dispersive effects
  are also important. These conditions are easy to satisfy in the context of cavity QED
  systems. Clearly if the spectral width of the vacuum is very large as in free space then
  one would not expect any significant change in branching ratios. Although the
  results
  that we present are specifically in the context of QED, they can be generally
  applicable to a much wider class of systems. For example we can consider the interaction
  of any system with a structured reservoir of finite width \cite{sten,zol,gs,kur}.
  Further the analysis would also
  be applicable to nano environments which lead to significant spectral modifications \cite{gsa2000}.\\

\begin{figure}[!h]
\begin{center}
\scalebox{0.65}{\includegraphics{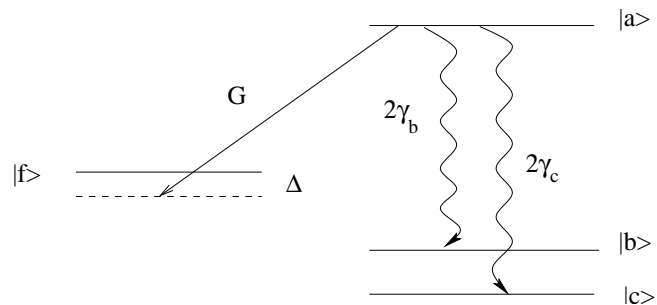}}
 \caption{Schematic diagram of a four-level atom. }
\end{center}
\end{figure}
  We start by illustrating the nature of the problem in the context of a simple four level
  model shown schematically in the Fig.1. Let us consider the decay of the excited
  state $|a\rangle$ to two lower levels $|b\rangle$ and $|c\rangle$. Let us also assume
   that the excited state $|a\rangle$ is connected to another level $|f\rangle$ by a laser
   field with Rabi frequency 2G. We can write all the density matrix equations for such a
   system clearly the population of the states $|b\rangle$ and $|c\rangle$ changes
   according to
\begin{equation}
\dot{\rho}_{bb} = 2\gamma_{b}\rho_{aa} \qquad , \qquad \dot{\rho}_{cc} = 2\gamma_{c}\rho_{aa}
\end{equation}
and thus
\begin{equation}
\frac{\rho_{bb}(t\rightarrow\infty)}{\rho_{cc}(t\rightarrow\infty)} = \frac{\gamma_{b}}{\gamma_{c}}
\end{equation}
The ratio is independent of the external field applied on the
transition $|a\rangle\rightarrow|f\rangle$. However the spectral
distribution itself would depend on the external field. We would
see later that result of Eq.(2) is intimately connected to the
Markov approximation, which is used in deriving Eq.(1). The
validity of the Markov approximation requires that the spectral
width of vacuum be much greater than the Rabi frequency of the
field that drives the transition $|a\rangle\rightarrow|f\rangle$.
This condition is satisfied in free space. Therefore the branching
ratios can be affected by considering vacuum whose width is
comparable with the applied external field or even less.
 This is also very relevant to the question of the interaction of multilevel systems with
 engineered reservoirs \cite{sten,zol,gs,kur}. Further we know from the early work of
 Purcell \cite{pur}
 that the spontaneous emission in a cavity is considerably modified \cite{haro} because
 the spectral width of the available mode is much smaller than in free space. Thus the question
 of the modification of branching ratios can be settled by considering emission in a cavity.\\

 To be specific we consider the case of Rydberg atom in a cavity which has a frequency
 $\omega_{c}$ which we can tune some where between the levels $|b\rangle$ and $|c\rangle$.
 The transition $|a\rangle\rightarrow|f\rangle$ is not resonant with the cavity and is driven by
 the laser field of frequency $\omega_{l}$. We now present a
 first principle calculation of the branching ratios. The Hamiltonian for the system in the interaction
 picture is given by,
 \begin{equation}
\eqalign{\mathbf{\hat{H}}& = G |a\rangle\langle f| e^{-i\Delta t}
+ \sum_{\omega} g_{b\omega}|a\rangle\langle b|
e^{-i\omega_{c}t+i\omega_{ab}t}\hat{a}_{\omega} + \nonumber\cr &
\sum_{\omega}g_{c\omega}|a\rangle\langle c|
 e^{-i\omega_{c}t+i\omega_{ac}t}\hat{a}_{\omega}+ h.c.\;,
 \qquad\Delta  =  \omega_{l} - \omega_{af}\;.}
 \end{equation}
 Here $\omega_{ab}$ and $\omega_{ac}$ represent, respectively, the frequencies of the two
 transitions. The cavity field is represented by the annihilation and creation operators $\hat{a}_
 {\omega}$,
$\hat{a}^{\dag}_{\omega}$. The coupling constant are denoted by $G
, g_{b\omega} , g_{c\omega}$. The sum over $\omega$ in Eq.(3)
would be converted into an inetgral
 over the spectral width of the single mode cavity. The wavefunction
of the system of the cavity field and the atom can be written as,
\begin{equation}
|\psi\rangle  =  \alpha|a,0\rangle + \beta|f,0\rangle
+\sum_{\omega}b_{\omega}|b,\omega\rangle +
\sum_{\omega}c_{\omega}|c,\omega\rangle.
\end{equation}
Here $\omega$ denotes the state of the cavity with one photon at
the frequency $\omega$. Various amplitudes can be obtained by
substituting Eqs.(3) and (4) in the Schr$\ddot{o}$dinger equation.
We work with Laplace transforms. The transform $\hat{\alpha}(z)$
of the excited state amplitude $\alpha(t)$ is given by,

\begin{equation}
\{
z+\frac{G^2}{z+i\Delta}+\sum_{\omega}\frac{|g_{b\omega}|^2}{(z+i\omega-i\omega_{ab})}
+\sum_{\omega} \frac{|g_{c\omega}|^2}{(z+i\omega-i\omega_{ac})}\}
\hat{\alpha}(z) = 1\;.
\end{equation}
Further the amplitudes of the final states $|b\rangle$ and $|c\rangle$ are found to be
\begin{equation}
\hat{b}_{\omega}(z) =
-i\frac{g^{\ast}_{b\omega}}{z}\hat{\alpha}(z+i\omega_{ab}-i\omega).
\end{equation}
\begin{equation}
\hat{c}_{\omega}(z) =
-i\frac{g^{\ast}_{c\omega}}{z}\hat{\alpha}(z+i\omega_{ac}-i\omega).
\end{equation}
We now convert the sums in Eq.(5) to integrals using,
\begin{equation}
\eqalign{\sum_{\omega}\frac{|g_{b\omega}|^2}{(z+i\omega-i\omega_{ab})}&\rightarrow\int{d\omega[\frac{
\kappa/\pi}{(\omega-\omega_{c})^2 +\kappa^2}]}
\frac{|g_{b}|^2}{(z+i\omega-i\omega_{ab})}\nonumber\cr & =
\frac{|g_{b}|^2}{(z+\kappa-i\delta_{b})}\;.}
\end{equation}
where $\delta_{b} = \omega_{ab}-\omega_{c}$. Finally one can prove
that the populations in the states $|b\rangle$ and $|c\rangle$
would be given by
\begin{equation}
P_{i} = |g_{i}|^2\int d\omega
\frac{\kappa/\pi}{(\kappa^2+\omega^2)}
|\hat{\alpha}(-i(\omega-\delta_{i}))|^2,\qquad i = b,c\;.
\end{equation}
 Note that we can identify 2$|g_{i}|^2/\kappa$ with the decay
2$\gamma_{i}$ of the state $|a\rangle$ to the state $|i\rangle$ in
a resonant cavity. This is the constant first calculated by
Purcell \cite{pur} and observed later by Goy. et. al. \cite{haro}.
The coefficient $\alpha$ given by Eq.(5) can now be written in a
more transparent form,
\begin{equation}
\hat{\alpha}(z) =
\{z+\frac{G^2}{z+i\Delta}+\frac{\kappa\gamma_{b}}{z+\kappa-i\delta_{b}}
+\frac{\kappa\gamma_{c}}{z+\kappa-i\delta_{c}}\}^{-1}\;.
\end{equation}

where $\Delta$ is the detuning of the coherent drive as shown in
the Fig.(1).The results given by Eq.(9) and Eq.(10) are the basic
results of this paper. These are exact-no approximation on the
coupling constant has been made. Similarly no approximation on the
strength of the coherent drive field has been used. Further all
dispersive effects are included through the complex Lorentzians in
Eq.(10) and therefore no Markov approximation is used. It may be
noted that the poles in $\hat{\alpha}(z)$ leads to spectral
modifications due to both coherent drive as well as due to strong
coupling effects \cite{wal,har,walt,kim}. The exact location of
such poles would depend on various detunings; field strength and
the coupling constants g. We do not discuss the issue of spectral
modifications in this paper.\\
\begin{figure}[!h]
\begin{center}
\scalebox{0.50}{\includegraphics{fig2.eps}}
 \caption{The populations $R_{b},R_{c}$ defined
 by $R_{b}=P_{b,\gamma_{b}=1,\gamma_{c}=1}/P_{b,\gamma_{b}=1,\gamma_{c}=0}$ ,
 $R_{c}=P_{c,\gamma_{b}=1,\gamma_{c}=1}/P_{c,\gamma_{b}=0,\gamma_{c}=1}$
 plotted as a
 function of detuning $\delta =(\omega_{ab}+\omega_{ac})-2\omega_{c}$ of the cavity in absence
 of any driving field. Here
 $\omega_{bc}$ is the separation between
 the states $|b\rangle$ and $|c\rangle$. All
 parameters are normalized with respect to $\kappa$. }
\end{center}
\end{figure}
\begin{figure}[!h]
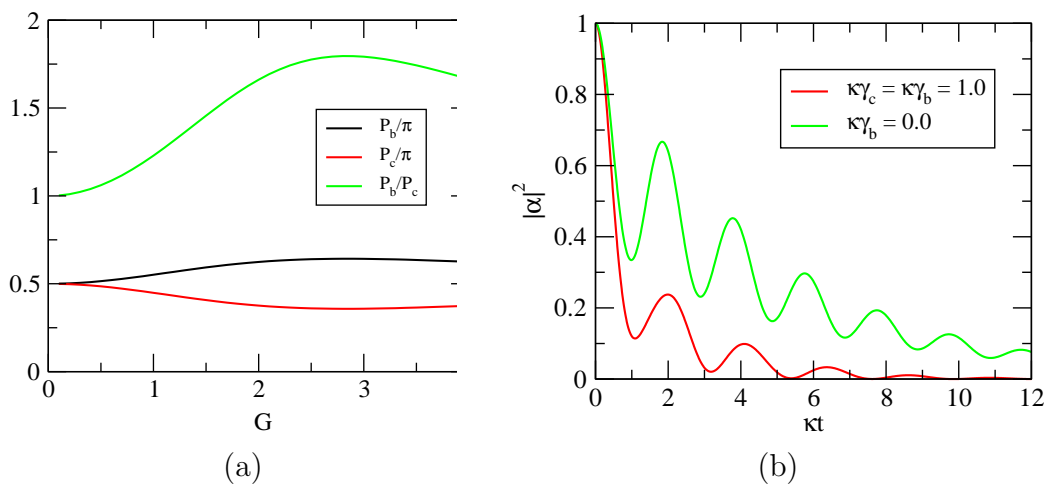

\begin{center}
\begin{tabular}{cc}
\scalebox{0.55}{\includegraphics{fig3.eps}}&
\scalebox{0.57}{\includegraphics{fig7.eps}}\\
(a)&(b)
\end{tabular}
\caption{(a) Populations of state
 $|b\rangle$ , $|c\rangle$ and the branching
 ratio $P_{b}/P_{c}$ as a function of the driving field, for cavity detuning
of $\delta_{b} = -\delta_{c} = 2.0$ and fixed drive detuning of
$\Delta = 2.0$. (b) Comparison between the
 population of the excited state as a function of time in the case of both lower states
 available and only one lower state available ($\gamma_{b}$ = 0) for G = 1.0,
$\delta_{c} = -2.0$ and $\Delta = 2.0$. The All
 parameters are normalized with respect to $\kappa$.}
\end{center}
\end{figure}

\begin{figure}[!h]
\begin{center}
\scalebox{0.50}{\includegraphics{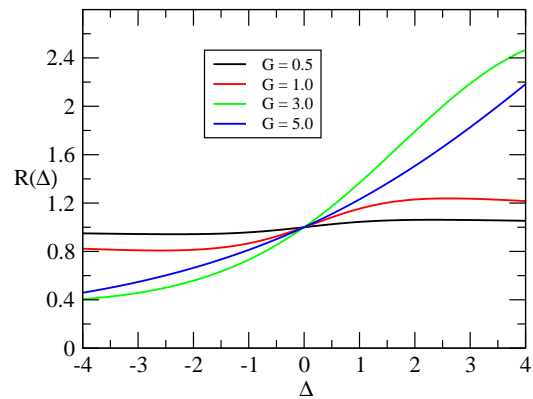}}
 \caption{Branching ratios $R = P_{b}/P_{c}$ plotted as a function of
 the detuning $\Delta$ of the coherent drive for
 different values of the driving field. The
 cavity detuning is kept fixed at $\delta_{b} = -\delta_{c}
 = 2.0.$ All parameters
 are normalized with respect to $\kappa$. }
\end{center}
\end{figure}

\begin{figure}[!h]
\begin{center}
\scalebox{0.50}{\includegraphics{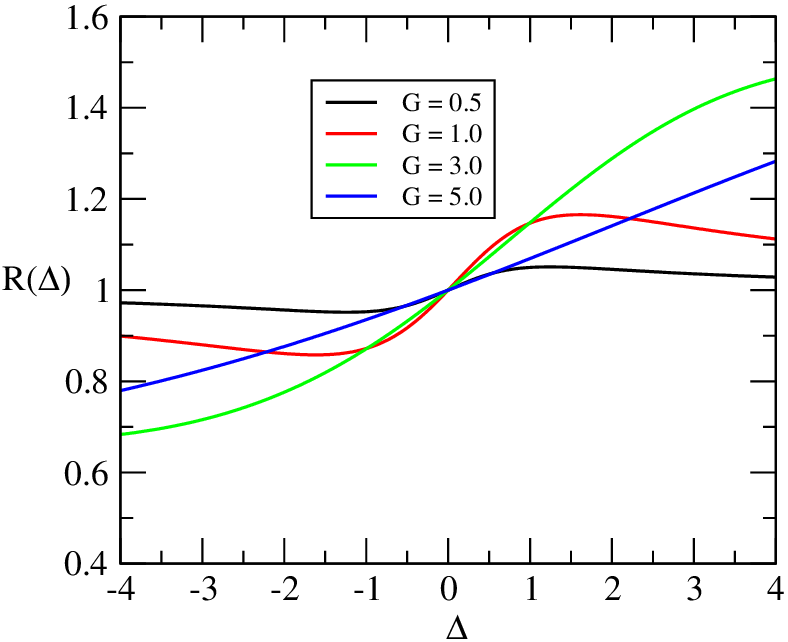}}
 \caption{Branching ratios $R = P_{b}/P_{c}$ plotted as a function of
 the detuning of the coherent drive as in Fig.4
 but now for a cavity detuning of $\delta_{b} = -\delta_{c}
 = 0.5.$ All parameters
 are normalized with respect to $\kappa$. }
\end{center}
\end{figure}

The branching ratio is given by $P_{b}/P_{c}$. In order to
highlight the effect of the external fields on the branching
ratios we consider some special cases first. Let us assume that
$\gamma_{b} =\gamma_{c} = 1$. Further assuming $\delta_{b} =
-\delta_{c}$ and $\Delta = 0$ one can prove that $P_{b} = P_{c}$.
Thus we do not find any dependence of the branching ratio on the
coherent drive. In order to obtain a asymmetric branching ratios
we can consider a cavity which is asymmetrically detuned from the
two transition frequencies. Even in the absence of any coherent
drive the asymmetric tuning can lead to $P_{b} \neq P_{c}$. We
show this in Fig.2 for a fixed separation between the states
$|b\rangle$ and $|c\rangle$ and for varying tuning of the cavity.
The effect gets more pronounced as the separation between two
states increases. Thus in order
 to obtain specifically the effect of the driving field we consider the symmetric
 situation $\delta_{b} =- \delta_{c}$. In Fig.3 we show how the branching ratio
 depends on the strength of the coherent drive. Note that $G/\kappa >
 1$ corresponds to the case when the Rabi frequency of the external
field exceeds the spectral width of the cavity vacuum. This is
typical of the situation when Markov approximation does not hold.
The effect of overlapping resonances and quantum interferences is
also very much evident in the time dependence of the population of
the excited state as shown in Fig. 3(b). In Figs. 4 and 5 we show
how the branching ratios change with the change in the detuning of
the coherent drive. The Fig.5 is for the case when the two
transitions $|a\rangle \leftrightarrow |b\rangle$ ; $|a\rangle
\leftrightarrow |c\rangle$ are within the
spectral width of the cavity. Thus in conclusion we have shown
how branching ratios can depend on external electromagnetic fields.
Our calculations show that the effects are especially pronounced if
 the vacuum of the electromagnetic field has a bandwidth comparable
 to the strength of the field. These kind of fundamental modifications
are expected to occur generally in any system-bath interaction.\\
\section *{Acknowledgement}
GSA thanks M. O. Scully and S. Suckewer for many discussions over
the years on the subject of branching ratios.

\end{document}